\documentclass{aa}  

\usepackage{graphicx}
\usepackage{hyperref}
\usepackage{amsmath}
\usepackage[T1]{fontenc}
\usepackage[utf8]{inputenc}
\usepackage{amssymb} 
\usepackage{physics}
\usepackage{subcaption}
\usepackage{txfonts}
\usepackage{gensymb}
\usepackage[dvipsnames]{xcolor}
\usepackage{orcidlink}

\usepackage[]{natbib}
\usepackage{hyperref}
\bibpunct{(}{)}{;}{a}{}{,}

\DeclareUnicodeCharacter{2212}{-}

\begin{document}

   \title{The three-dimensional shapes of the galaxy cluster intracluster medium in eRASS1}

   \author{S. Kulkarni\orcidlink{https://orcid.org/0009-0000-9406-7721}
          \inst{1,2}
          \and
          J. S. Sanders \orcidlink{https://orcid.org/0000-0003-2189-4501}\inst{1}
          \and
          E. Bulbul\orcidlink{https://orcid.org/0000-0002-7619-5399}\inst{1}
          \and
          A. Liu\inst{1,3}
          \and
          M. E. Ramos-Ceja\orcidlink{orcid:0000-0002-9117-3251}\inst{1}
          }
   \institute{Max-Planck-Institut f\"{u}r Extraterrestrische Physik, 
   Gie\ss enbachstraße 1, 85748 Garching, Germany
         \and
             Ludwig-Maximilians-Universit\"at  M\"unchen, Scheinerstr. 1, 81679 Munich, Germany,
        \and 
            Institute for Frontiers in Astronomy and Astrophysics, Beijing Normal University, Beijing 102206, China}
             
   \date{Received --; accepted --}

  \abstract
 
   {The three-dimensional shapes of clusters are important for understanding the astrophysics of the clusters and as a probe in cosmological studies.}
   {We estimate the most probable three-dimensional shape of galaxy clusters in the first eROSITA All-sky survey eRASS1 using stereology. Our sample is the largest well-defined sample of clusters, and the most probable shape estimated using our method can be used as a prior for cluster shape models in cosmological, cluster, and weak lensing studies.}
   {The first all-sky survey with SRG (Spectrum Roentgen Gamma)/ eROSITA resulted in a sample of $\sim$12,000 optically confirmed galaxy groups and clusters. We used a well-defined subsample of 3254 clusters from the eRASS1 survey and estimated the most probable shape of the clusters by constraining the probability density function (PDF) of the ellipticity of the clusters. We simulated the projected appearance of clusters with a distribution of three-dimensional shapes (prolate and oblate) and obtained the distribution of their ellipticity. This distribution was then compared with the measured distribution of ellipticities from the eRASS1 cluster sample to infer the three-dimensional shapes consistent with the data. We used Monte Carlo methods to estimate the most probable axial ratios $l,w$, where $l\equiv L/T$,$w\equiv W/T$, and $L, W, T$ are major, intermediate, and minor axes of the cluster. We did not require any additional probe (optical, SZ, etc.) to constrain the probable shape of the clusters.}
   {We describe the ellipticity PDF of the eRASS1 clusters with a normal distribution $(\mathcal{N})$ with a mean ($\mu$) and a standard deviation ($\sigma$), $\mathcal{N}(\mu,\sigma) = (0.79,0.25)$. The most probable shape of the clusters in our eRASS1 subsample is estimated to be $(l,w)=(1.51 \pm 0.27,1.17 \pm 0.27)$, with prolate shapes being preferred over oblate shapes.}
   {}


   \maketitle
%
\section{Introduction}

Galaxy clusters are the largest gravitationally bound objects in the Universe. They form as a result of the gravitational collapse of primordial matter overdensity and lie at the intersection of cosmic filaments. Galaxy clusters are laboratories for studying the dynamics between dark matter and its baryonic components (e.g., the intracluster medium (ICM), and galaxies). They are important probes used for constraining cosmological parameters and for testing theories of gravity \citep[e.g.][]{Wojtak_2011, Rosselli_2023, Ghirardini_2024, Seppi_2024srgerosita, Artis_2024, Artis_2025}. The mass of clusters used in the halo mass function is used as a cosmological probe. A method of estimating the mass of clusters assumes the hydrostatic equilibrium (HSE) between the dark matter halo and the ICM gas. A particular cluster geometry (often spherical) is assumed in HSE assumptions, which introduces biases in mass estimation studies. This mass bias can be reduced by assuming the nonspherical shapes of the clusters\citep{Jetzer_2002, Lee_2003, Gavazzi_2005, Kravtsov_2012, Limousin_2013, Ettori_2013, Parekh_2015, Chiu_2018, Sanders_2018, Yuan_2020, Ghirardini_2022, Herbonnet_2022, Veronesi_2025}. Considering the significance of clusters as cosmological probes, their fundamental properties, such as shape, mass, temperature, and the physical processes influencing clusters, need to be studied in detail.

There is extensive historical literature suggesting that clusters are nonspherical, including early observations from galaxy distributions \citep[e.g.][]{Rood_1972, MacGillivray_1976, Carter_1980}, X-ray imaging \citep[e.g.][]{Hirayama_1978, Strimpel_1979, Fabricant_1984, Buote_1992, Buote_1995, Limousin_2013}, and weak-lensing studies \citep[e.g.][]{Oguri_2010, Sereno_2011, Oguri_2012, Chiu_2018}. On the theoretical side, various studies \citep[such as][]{Zeldovich_1970, White_1979} suggest clusters' nonspherical shapes. The collapsed object evolves such that the length of the shortest axis becomes zero, forming a pancake-like structure. Nonspherical shapes of clusters can be easily explained if the initial density perturbations were anisotropic. Their evolution via accretion through the filamentary structure over time amplifies the initial anisotropies \citep{White_1979, Eisenstein_1995}. Cosmological N-body simulation studies \citep[e.g.][]{Warren_1992, Shaw_2006, Schneider_2012, Lau_2020} reported that the clusters have nonspherical shapes, and a preference for prolate shapes. Despite the considerable evidence of nonspherical cluster shapes, the approximation of spherical symmetry is common.

The shape of the components of galaxy clusters, such as the ICM, dark matter halo, and brightest cluster galaxy (BCG), can tell us about the current dynamic state of the clusters along with their dynamic history. The offset between the position of the BCG in the cluster and the centroid of the ICM shape and the centroid of the DM halo might suggest a recent merger or disturbance due to internal processes. There appears to be a strong correlation between the shape alignments of the BCG and the ICM in clusters \citet{Martin_2010, Ragone_2020} and \citet{Gassis_2023}. Previous studies, such as \citet{Chambers_2002} and \citet{Ho_2006}, have suggested the use of shapes and their alignment properties as a cosmological probe.

Estimating the correct shapes of the clusters is important for cluster and cosmological studies. Yet, estimating the three-dimensional (3D) shapes is difficult primarily due to projection biases. These projection biases can be overcome by de-projecting the two-dimensional (2D) cluster shapes to 3D cluster shapes. To reconstruct the 3D shape using de-projection methods, multiple probes, such as optical, X-ray, Sunyaev–Zeldovich (SZ), or weak-lensing, are required \citep{Basilakos_2000,  Sereno_2007, Chakrabarty_2008}. While the availability of multi-probe datasets for all the detected clusters in eRASS1 is an issue, combining these datasets is another. These difficulties can be overcome by the method developed by \citet{Makarenko_2014} and \citet{Shankar_2021}, which uses a sample of clusters with a singular probe. Their method uses the probability density function (PDF) of the shape parameter to estimate the probable axis ratios by using Monte Carlo methods.

We applied their method to a subsample of 3254 clusters from the eRASS1 catalogue \citep{Bulbul_2024, Kluge_2024}. The first eROSITA all-sky survey (eRASS1) covers $13,116$ deg$^2$ of the sky. eRASS1 provides the largest X-ray sample of clusters containing over 12,000 optically confirmed galaxy clusters and groups, over $0<z<1.3$. A statistically well-defined, purer sample of clusters containing $\sim$5200 clusters with optical follow-up and matched with other X-ray survey samples is available for cosmological studies \citep{Ghirardini_2024, Artis_2024, Artis_2025}. Morphological parameters and other X-ray properties of the clusters are also available for these clusters \citep{Sanders_2025}. We selected the brightest and largest clusters from the 5200 clusters by applying constraints on the detection likelihood and extent likelihood \citep{Brunner_2022, Liu_2022, Merloni_2024}. Our sample of clusters contains 3254 clusters. This is the statistically largest and purest cluster sample to which we apply this method. This paper is structured as follows. In section \ref{Sec2}, we discuss the methodology, in section \ref{Sec4}, we discuss the results, and in section \ref{Sec5}, we conclude this study.
\section{Methodology}
\label{Sec2}
The 3D structure of the clusters cannot be estimated using a single observational probe (e.g., optical, X-ray, or SZ datasets); we have to gather information from the 2D projection of the cluster along the line of sight. As it is difficult to know how the cluster is oriented in 3D space, 2D projections of clusters lead to uncertainties in shape estimations. Hence, any projection of the cluster is probabilistic. To obtain the information about the shape of the cluster, we performed a statistical analysis \citep{Vinci_2014}. The study of estimating 3D shapes by using 2D cross sections is called stereology. Stereology combines the information of projections of known geometries and statistics to estimate morphological information, such as shape. Stereology allows us to reconstruct the probable 3D structure using available 2D cross sections of the object, solving the de-projection problem. This method has already been applied by \citet{Vinci_2014} to estimate the probabilistic shapes of galaxies.
\subsection{Shape estimation using stereology}
\label{Sec:2.1}
Stereological methods can be used to describe shape, provided the shape description parameter is a Minkowski functional \citep{Bharadwaj_2000, Makarenko_2014, Shankar_2021}. Since ellipticity as a shape descriptor is a Minkowski functional, the stereological method is applicable \citep[and references therein]{Makarenko_2014, Shankar_2021}. We aim to determine the most probable shape of the clusters in our cluster sample. The 3D geometry of the cluster can be expressed by its length ($L$), width ($W$), and thickness ($T$). Keeping in mind the projection effects, it is easier to determine the ratios $l\equiv L/T$ and $w\equiv W/T$ instead. For our study, we define $T=1$, as we are only interested in finding the shape, not the size of the clusters. 
We estimated the ellipticity of projections of randomly oriented clusters with known ($l,w$) and constructed the conditional probability density function (PDF) ($\mathcal{P}(e|l,w)$).

We estimated the ellipticity of the 3254 clusters in our sample and constructed an observational PDF ($\mathcal{P}_{obs}(e)$). We aim to find the correct combination of ratios ($l,w$) that reconstruct the observation PDF ($\mathcal{P}_{obs}(e)$). To find the correct combination of ($l,w$) that reconstructs the observational PDF, we generated a sample of ratios from a prior distribution and used Monte-Carlo methods to estimate the ratios ($l,w$).
\begin{figure*}[h!]
    \centering
    \includegraphics[width=0.9\linewidth]{ 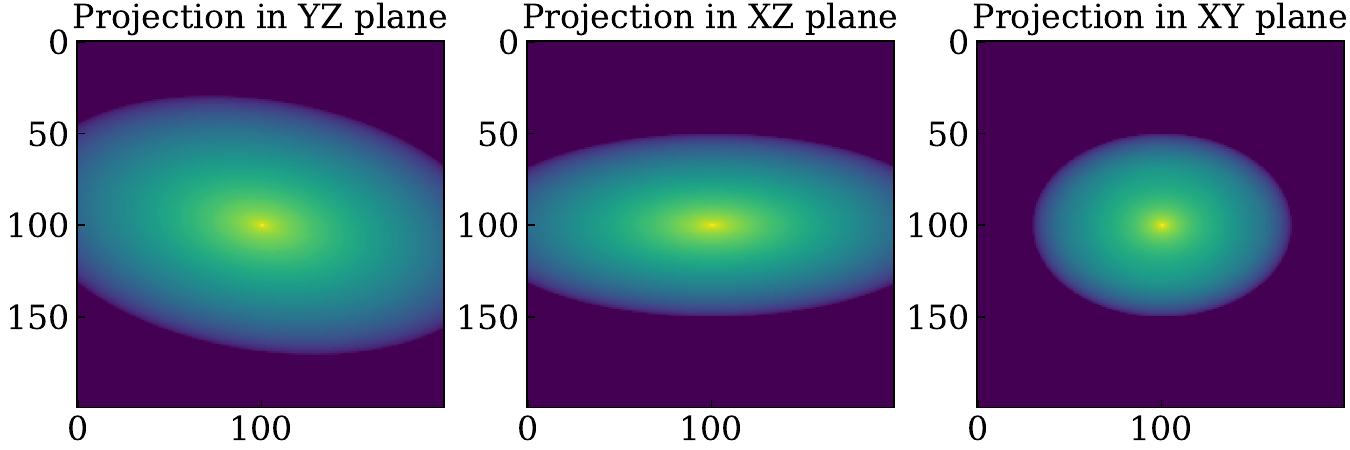}
    \caption{Projection effects for the Vikhilinin model surface brightness profiles (see section \ref{Sec:3.1}). The projection of the cluster with length axis $L: W: T=2:1.5:1$, and Euler rotation of $30\degree$ along the X axis, is not similar along different axes.}
    \label{fig:1.1}
\end{figure*}
\subsection{Projection effects and biases}
\label{Sec:2.2}
To explain the projectional effects, figure \ref{fig:1.1} shows an example projection of a cluster in three dimensions with axis lengths $L:W: T=2:1.5:1$, along the $X$, $Y$, and $Z$ axes and a Euler rotation of $30\degree$ about the $X$ axis. The projection along the different axes is not similar; hence, the estimated projected shape of the cluster will depend on the observer's position. To estimate the 3D structure of the cluster, one needs to deproject the 2D surface brightness maps into 3D data while taking into consideration:
\begin{enumerate}
    \item The coordinate system of the observer; and
    \item The geometry of the cluster. For de-projection, one needs to assume a shape and the orientation of the cluster.
\end{enumerate}
The orientation of the cluster can be fixed by analyzing how the de-projections of clusters with random cluster orientations vary. \citet{Chakrabarty_2008} suggested that these relations are unique, but some studies suggest otherwise \citep{Plionis_1991}. 

Following our definition of the ratios, $l\approx w\approx 1$ describes spherical shapes, while $l\geq w \approx 1$ describes prolate shapes. Clusters with extreme ratio values, i.e., $l >> w \geq 1$, describe ellipsoidal or triaxial shapes. However, clusters in reality do not have completely spherical shapes, and nor do they have extremely ellipsoidal or triaxial shapes. Hence, we want our estimates to be realistic, which is achieved by constraining our values to $e \in (0.2,0.8)$, where $e$ is the ellipticity parameter defined as ($e\equiv b/a$), the ratio of the semiminor axis ($b$) to the semimajor axis ($a$). Selecting realistic ellipticity values does not affect the selection \citep[see ][]{Sanders_2025}.

We find our method to have potential biases for the following reasons. 
\begin{enumerate}
    \item Selection of the cluster sample: As the method finds the most probable shape of the clusters in a given sample, the results depend on the choice of the cluster sample. The results for a sample of clusters with spheroidal geometries and non-spheroidal geometries differ significantly. For example, X-ray selected clusters are typically more spherical than weak-lensing selected clusters. Application of the method to these two samples might have slightly different results. However, one can use this to one's advantage by carefully choosing the cluster sample. Moreover, the cluster sample should be statistically complete and uncontaminated. 
    \item Selection of prior distributions to simulate the shapes: The prior distribution $\mathcal{U}(a,b)$ or $\mathcal{N}(\mu,\sigma)$, which we draw the sample of ratios $(l,w)$ from and use to construct the conditional PDF $\mathcal{P}(e|l,w)$, must be such that shapes represented by the observational PDF, $\mathcal{P}_{obs}(e)$, are contained in the sample. 
    \item Intrinsic biases due to the projection effects: These effects are intrinsic to any study. The shape estimated is the shape estimated from the projection toward the observer. This shape might be the intrinsic shape of the cluster, but to prove that, we would need the projection of the cluster from different points in space, which is not possible. Since the orientation of the cluster is unknown, this may introduce some bias toward certain shapes.
    \item Biases due to stereological techniques: These effects will arise from the inaccurate or unavailability of information about the geometry of an object. Stereological techniques work well with well-studied geometries such as spheres, ellipsoids, or pyramids. From the observations, we know that cluster geometries can depart from ellipsoidal shapes as clusters are complex structures with multiple processes affecting the morphology of clusters. These biases can be reduced by developing stereological techniques for complex shapes. We would like to emphasize that stereological bias does not apply to our study, as spherical and ellipsoidal geometries are well studied.
\end{enumerate}
While the selection bias and the prior distribution bias can be reduced, the intrinsic bias due to projection effects cannot be completely removed.
\subsection{Ellipticity calculation}
\label{Sec:2.3}
Ellipticity is a non-negative number that describes the shape of the conic section. For our study, we define the ellipticity as 
\begin{equation}
    e = b/a,
\label{eqn:1.1}
\end{equation}
where $b$ is the semiminor axis, and $a$ is the semimajor axis. According to our definition, a line has $0$ ellipticity, and an ellipse has ellipticity $0 < e \leq 1$, and $e=1$ describes a circle.\footnote{Some other authors use definition of ellipticity, $\epsilon = 1-b/a$, so a circle has $\epsilon=0$ and a line has $\epsilon=1$. We took this into account in our analysis and converted the values according to our definition.}
To make our analysis computationally faster and to reduce biases toward certain cluster models, we did not use any particular cluster profile model. Instead, we simulated a 3D cluster by choosing three axis lengths ($L, W, T$), and took the projection along a random normal vector. This projection along the normal vector, in two dimensions, is described by an ellipse. This simulates the projection of a 3D cluster (at the boundary of the cluster), with length $(L)$, width $(W)$, and thickness $(T)$, randomly oriented in the 3D space. 
An ellipse can describe any such projection of a triaxial cluster. We calculated the ellipticity of the ellipses by using the affine transformation property for the ellipse. While the affine transformation changes the Euclidean lengths, this is not a concern for us as we are only interested in the ratio of the axes and not the actual lengths. Any ellipse is an affine image of the unit circle $x^2 + y^2 = 1$. This transformation can be defined as
\begin{equation}
    \vec{x} \rightarrow \vec{f_0} + A \vec{x}, 
\end{equation}
where A is the transformation matrix and $\vec{f_0}$ is an arbitrary vector. (If the projection plane is the same as the XY plane, $\vec{f_0}$ is the zero vector.) If $\vec{f_1}$, $\Vec{f_2}$ are the column vectors of A, then the circle ($\cos{t},\sin{t}$), $0 \leq t \leq 2 \pi$ is mapped to an ellipse, i.e., 
\begin{equation}
    \vec{x} = \vec{f_0} +  \vec{f_1}\cos{t} + \vec{f_2}\sin{t},
\end{equation}
where $\vec{f_0}$ is the center and $\vec{f_1}$, $\vec{f_2}$ are the conjugate axes. To calculate the ellipticity, we estimated $\vec{f_1}$, $\vec{f_2}$. We simulated the line-of-sight projections of the ellipsoidal surface brightness map. We then used an affine transformation to find the matrix A. The transformation matrix then gives us $\vec{f_1}$, $\vec{f_2}$. Then we used the vector with the minimum length as the semiminor axis ($b$) and the vector with maximum length as the semimajor axis ($a$), and calculated the ellipticity using the equation \ref{eqn:1.1}.

We can generate a probability distribution of ellipticity for the clusters in our sample, $\mathcal{P}_{obs}(e)$. This PDF can be generated in two ways:
\begin{enumerate}
    \item If we observe a cluster with a given value of $l$,$w$ from random directions and with random orientations, and calculate the ellipticity for all of the projections and construct the PDF, or
    \item If we observe a large number of different clusters, each with the same value of $l$,$w$, calculate the ellipticity for them and construct the PDF.
\end{enumerate}
These two PDFs are then identical \citep{Makarenko_2014, Shankar_2021}. Our observational PDF $\mathcal{P}_{obs}(e)$ is constructed using the second process. We can construct the same PDF using the first process with the known values of $l, w$. This can be written as follows:
\begin{equation}
    \mathcal{P}_{obs}(e) = \int dl dw \mathcal{P}(e,l,w),
\end{equation}
\begin{equation}
    \mathcal{P}_{obs}(e) = \int dl dw \mathcal{P}(e|l,w)\mathcal{P}(l,w),
    \label{eqn:5}
\end{equation}
where $\mathcal{P}_{obs}(e)$ is the PDF constructed either by process 1 or 2. Since both PDFs are identical, we treat them the same. $\mathcal{P}(e|l,w)$ is called the conditional PDF, and $\mathcal{P}(l,w)$ is the PDF that contains information about the shape of the clusters. Observational PDF, $\mathcal{P}_{obs}(e)$, is a function of ellipticity that depends on the intrinsic shapes of the clusters in our sample. The conditional PDF changes as we vary the ratio pair ($l$,$w$). If we choose the sample such that $l >> w$, then the clusters will be more eccentric. If we choose ratios such that $l\approx \ w\approx 1$, the clusters should be more spherical. Conditional PDFs should reflect this.
\subsection{Conditional PDFs of ellipticity}
\label{Sec:2.4}
\begin{figure}[]
    \centering
    \includegraphics[width=\linewidth]{ 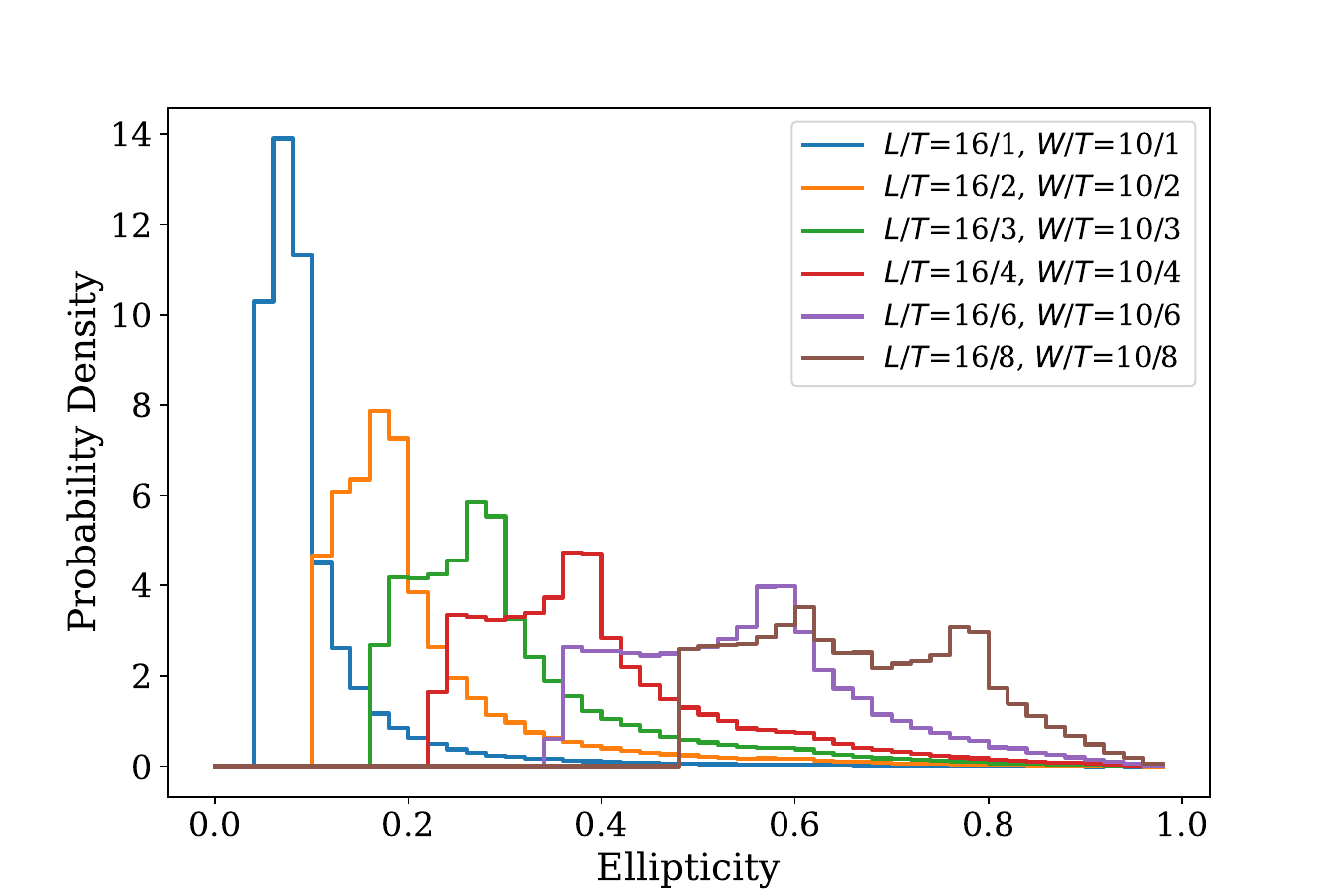}
    \caption{Conditional PDF for varying thickness with constant $L=16$ and $W=10$. We vary $T$ from 1 to 8. As the shape gets spherical, i.e., the ratios $l,w \rightarrow 1$, the conditional PDFs shift to the right.}
    \label{fig:3}
\end{figure}

We simulated the conditional PDFs by taking $\sim$150,000 random 2D projections of a cluster with known axis lengths $(L, W, T)$ of the ellipsoid, calculating the ellipticity of the 2D projections, and constructing the PDF.
\begin{figure}[]
    \centering
    \includegraphics[width=\linewidth]{ 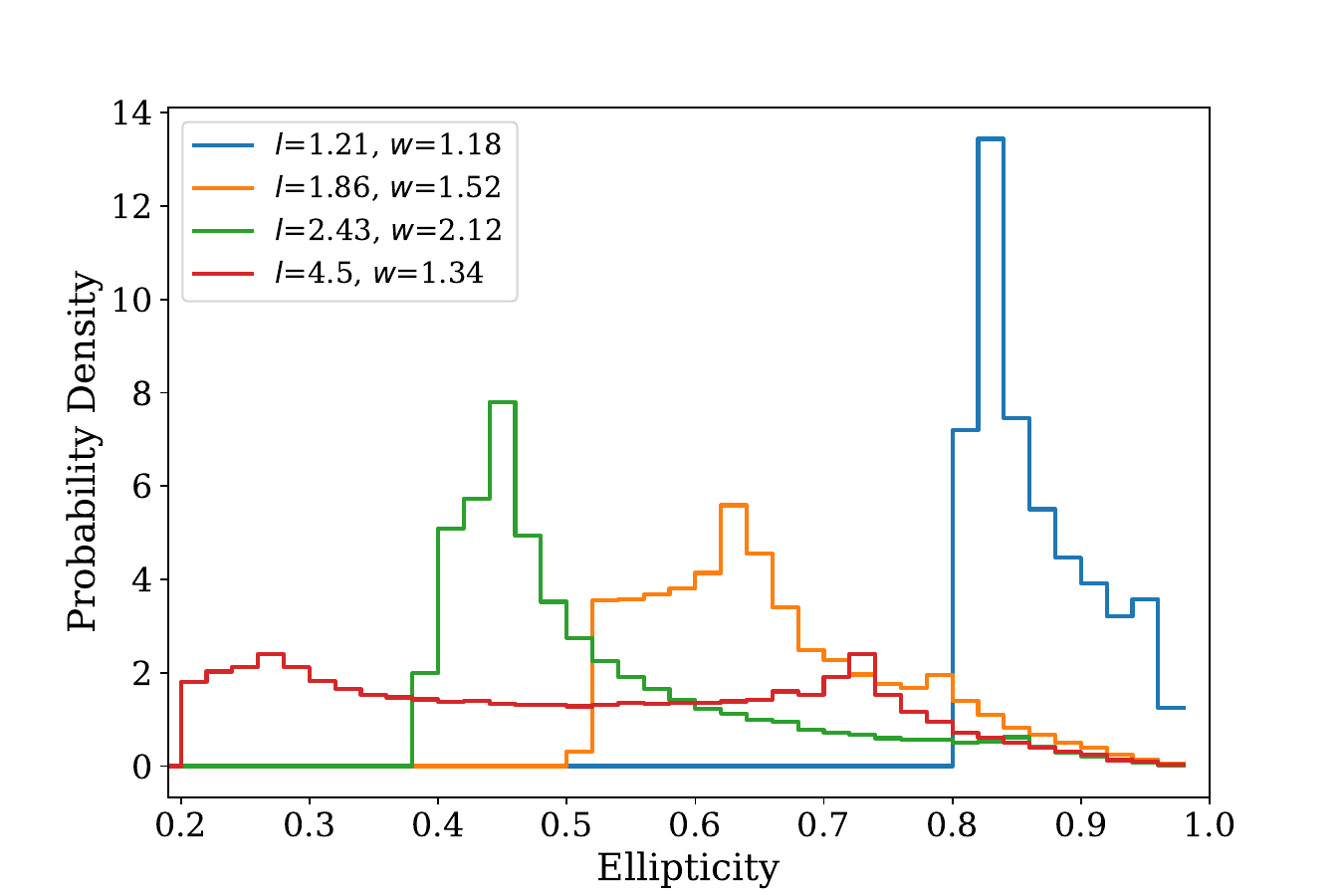}
    \caption{Conditional PDFs for known ratios $l,w$. The blue PDF (ratios $l = 1.22,w = 1.18$) peaks at ellipticity $e\sim0.86$, representing spheroidal shapes, while the green PDF ($l = 2.43,w = 2.12$) peaks at $e\sim0.47,$ representing ellipsoidal shapes. For extreme shapes ($l = 4.5,w = 1.34$), the probability density decreases, and we get a bimodal distribution which peaks at $e_{peak} \sim0.3$ and $e_{peak}\sim0.75$, due to projection effects.}
    \label{fig:2}
\end{figure}
In Fig. \ref{fig:2}, we plot the conditional PDFs for different values of ratios $l\equiv L/T,w\equiv W/T$. Most PDFs show a sharp peak and then decline. \
In Fig. \ref{fig:3}, we plot the PDF of the ellipticity of the cluster with constant length $(L=16)$ and width $(W=10)$ and vary the thickness. We observe a similar behavior to what we observed in Fig. \ref{fig:2}. We observe that as the values of both ratios ($l,w$) approach 1, the PDF shifts to the right (brown PDF), describing spherical shapes. 
From Figs. \ref{fig:2} and \ref{fig:3}, it can be seen that the conditional PDFs span the complete ellipticity space and for a given $l,w$ pair there is a conditional PDF, $\mathcal{P}(e|l,w)$, which is unique to the choice of $l$ and $w$ \citep{Makarenko_2014, Shankar_2021}. Conditional PDFs vary according to changes in $l$ and $w$. The peak value of the PDFs spans the parameter space, and we can generate a set of conditional PDFs that describe all shapes ranging from spherical to ellipsoidal geometries, extending to triaxial geometries. 

Taking another look at equation \ref{eqn:5}, out of the three terms, we know the conditional PDF, $\mathcal{P}(e|l,w)$, and the observational PDF, $\mathcal{P}_{obs}(e)$. The only unknown remaining is the shape PDF, $\mathcal{P}(l,w)$, which contains the information about the shape of the clusters \citep{Shankar_2021}. We aim to find this PDF. To make the model easier and computationally inexpensive, we can approximate the shape PDF ($\mathcal{P}(l,w)$) by using the Dirac-delta function, i.e.,
\begin{equation}
    \mathcal{P}(l,w) \approx \sum_{i=1}^{n} a_i \delta_{D}(l-l_i,w-w_i),
    \label{eqn:6}
\end{equation}
where $\delta_{D}$ is the Dirac delta function at $(l_i,w_i)$ in the $l,w$ plane. Here, $n$ is the number of $(l,w)$ pairs in the sample (i.e., sample size) and $a_i$ is the probability associated with the $i^{th}$ pair $(l_i,w_i)$, with the normalization condition for probability,
\begin{equation}
    \sum_{i=1}^{n} a_i = 1.
\end{equation}
Since we use discrete values for $l$,$w$ pair, approximating the shape PDF ($\mathcal{P}(l,w)$) with the Dirac delta function will only approximate the results \citep{Shankar_2021}. Instead of working with continuous variables ($l,w$), we substituted equation \ref{eqn:6} in the equation \ref{eqn:5} and used the translation properties of the Dirac function and integrated the Dirac-delta function over $l,w$ space, so that the equation reduces to
\begin{equation}
    \mathcal{P}_{obs}(e) \approx \sum_{i=1}^{n} a_i \mathcal{P}(e|l_i,w_i).
    \label{eqn:8}
\end{equation}
Now, our observational PDF can be approximated by finding the correct combinations of conditional PDFs for the $(l_i,w_i)$ pair and the probability, $a_i$, associated with it. 

After quantifying the observational PDF, our problem of finding the most probable shape of the clusters reduces to finding the model PDF, $\mathcal{P}_{model}(e|l_i,w_i)$, which is a combination of conditional PDFs associated with $n$ $(l,w)$ pairs drawn from a known distribution. Given that the model PDF fits the observational PDF ($\mathcal{P}_{obs}$) correctly, i.e.,
\begin{equation}
    \mathcal{P}_{obs}(e) \approx \mathcal{P}_{model},(e|l,w),
    \label{eqn:9}
\end{equation}
the most probable pair(s) $(l,w)$ from the sample of $n$ pairs can be found by estimating the probability, $a$, associated with the pairs \citep{Shankar_2021}.

We proceeded further by identifying the sample of ($l,w$) pairs that can reconstruct the observational PDF ($\mathcal{P}_{obs}$). Then, we used Monte Carlo estimation to narrow down our sample and find the most probable pair. 
To estimate the most probable pair(s) using Monte Carlo methods, we first drew a sample with $n$ pairs of $(l,w)$. The sample that correctly fits our observational data contains the most probable pair ($l,w$) \citep{Shankar_2021}. This can be done by generating random pairs of ($l,w$) from a prior probability distribution. For example, sample $n$ pairs of ($l,w$) from a normal distribution with mean ($\mu$) and standard deviation ($\sigma$). Then, we assigned equal probability to all the $n$ pairs in our randomly generated ($l,w$) sample. This means that all shapes from most eccentric to spherical are equiprobable, i.e.,
\begin{equation}
    a_i = 1/n,
\end{equation}
where $n$ is the number of pairs of ($l,w$) in the sample. We assigned equal probability to all the pairs, as we did not want to make any assumption about the shape a priori \citep{Shankar_2021}. Another reason for equiprobable shapes is the lack of known priors on shapes from the studies done so far.

Next, we used the chi-square statistic to find the best-fitting PDF. The statistic is defined as 
\begin{equation}
    \chi^2 = \sum_{j} \frac{[\mathcal{P}(e_j)-\mathcal{P}_{obs}(e_j)]^2}{\sigma_j^2}
    \label{eqn:11}
,\end{equation}
where $\mathcal{P}(e_j)$ is the counts in the $j^{th}$ bin of the constructed PDF, similarly for the $\mathcal{P}_{obs}(e_j)$, and $\sigma_j^2$ is the Poisson error calculated as $\sigma_j^2 = (\sqrt{n_{obs}(j)}/n_{total})*n_{bin}$, where $n_{obs}(j)$ are the counts in the $j^{th}$ bin, and $n_{total}$ are the total counts and $n_{bin}$ is the number of bins used \citep{Shankar_2021}.

To estimate the most probable ratio, we started by calculating the $\chi^2$ statistic between the model PDF (equation \ref{eqn:9}) and the observed PDF. Now, we want to reduce the chi-square value by removing unlikely distributions. The model PDF can be written as
\begin{equation}
    \mathcal{P}_{model}(e) = \sum_{i=1}^{n} a_i P(e|l_i,w_i). 
    \label{eqn:12}
\end{equation}
Since all the pairs have equal probability a priori,
\begin{equation}
    a_i = 1/n. 
\end{equation}
From these $n$ pairs, we randomly removed one pair ($l,w$) and summed the remaining conditional PDFs associated with $n-1$ pairs. This new model PDF is given by
\begin{equation}
    \mathcal{P}_{new}(e) = \sum_{i=1}^{n-1} a_i \mathcal{P}(e|l_i,w_i) 
,\end{equation}
with
\begin{equation}
 a_i = 1/n-1. 
\end{equation}
We estimated the $\chi^2$ between $\mathcal{P}_{new}(e)$ and observed PDF and called it $\chi_{new}^2$, i.e., 
\begin{equation}
    \chi_{new}^2 = \sum_{j} \frac{[P_{new}(e_j)-P_{obs}(e_j)]^2}{\sigma_j^2}
    \label{eqn:16}
,\end{equation}
where $\mathcal{P}_{new}(e_j)$ is the counts in the $j^{th}$ bin of the model PDF (with $n-1$ pairs), similarly for the $\mathcal{P}_{obs}(e_j)$, and $\sigma_j^2$ is the Poisson error calculated as before.

If: 
\begin{enumerate}
    \item $\chi_{new}^2$ $<$ $\chi^2$, the pair is not necessary to the model PDF and can be removed.
    \item $\chi_{new}^2$ $>$ $\chi^2$, the pair is essential to the PDF and can not be removed.
\end{enumerate}
As a point is removed, the probability of the remaining points increases \citep{Shankar_2021}. We updated the chi-square value as follows:
\begin{enumerate}
    \item If point is removed, $\chi^2$ = $\chi_{new}^2$ given by equation \ref{eqn:16}.
    \item If a point is not removed, $\chi^2$ is unchanged.
\end{enumerate}
We followed this procedure until $\chi_{new}^2 \approx \chi^2$, and it does not vary much. 

The new model PDF, which we are left with, can be written as
\begin{equation}
    \mathcal{P}_{final}(e) = \sum_{i=1}^{k} a_i P(e|l_i,w_i),
    \label{eqn:17}
\end{equation}
where $k < n$ and the probability of the $k^{th}$ pair is given by
\begin{equation*}
    a_i = 1/k.
\end{equation*}
Finally, we approximated the $l$,$w$ further by binning the $l$,$w$ space and estimated the probability for the bins. The density of points in a bin increases the probability for that bin. 
\section{Data analysis}
\subsection{Ellipticity PDF of clusters in eRASS1}
\label{Sec:3.1}
We used a subsample of 3254 clusters over the redshift range $0\leq z \leq1$, from $\sim 12,000$ eRASS1 clusters. We chose the brightest clusters from eRASS1 by selecting the clusters with \texttt{DET\_LIKE} $>$ 40 and \texttt{EXT\_LIKE} $>$ 6  \citep{Brunner_2022, Liu_2022, Merloni_2024, Bulbul_2024}, where DETLIKE is the detection likelihood and EXTLIKE is the extent likelihood. The morphological parameters were calculated for every cluster in our sample \citep{Sanders_2025}. First, the X-ray image was cleaned by removing the background exposure and point sources, and the unnecessary objects were masked. 

\begin{figure}
    \centering
    \includegraphics[width=\linewidth]{ 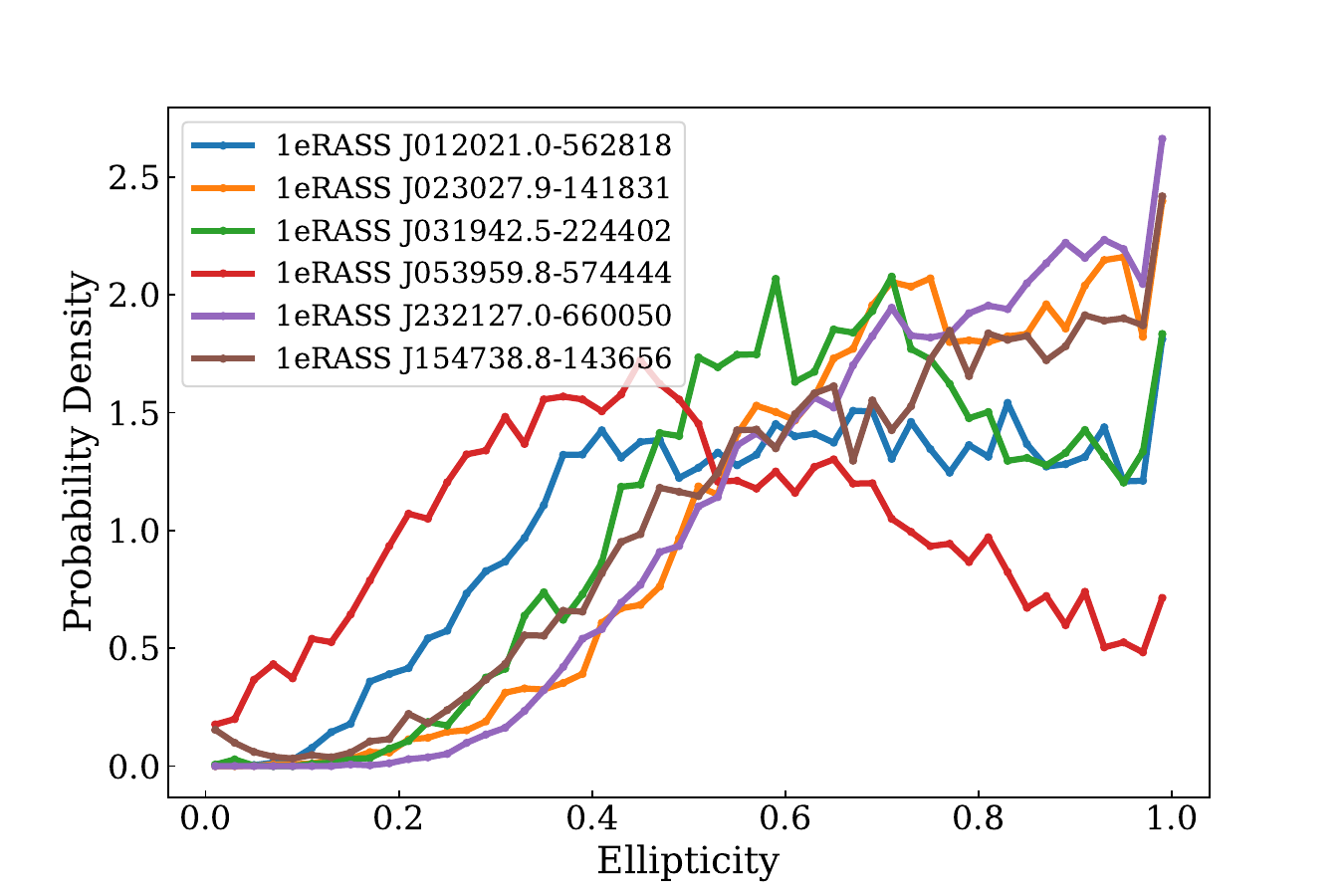}
    \caption{Posterior PDFs for some clusters in our sample. On the x axis, we plot the ellipticity, and on the y axis, we plot the probability density. The legend contains the name of the cluster.}
    \label{fig:4}
\end{figure}

The Vikhlinin emission model, with axial symmetry along the $z$ axis, was used to model the cluster's surface brightness \citep{betaprof_Vikhlinin_2006}. The surface brightness profile model (equation \ref{eqn:18}) was fit to the surface brightness map of the cluster, where $S(r)$ is the original radial profile constructed for the cluster, and $x$,$y$ are the 2D coordinates of the projected surface brightness maps. Ellipticity ($e$) and the angle ($\theta$) made by the cluster with the $X$ axis on the surface brightness map were varied to find the best-fit values. Markov chain Monte Carlo (MCMC) provides a posterior probability distribution \citep[see][]{Sanders_2025}: 
\begin{equation}
S'(x,y) = S(e[x \cos\theta_0 - y\sin\theta_0]^2 + e[x\sin\theta_0 - y\cos\theta_0]^2/e )^{1/2}
\label{eqn:18}
.\end{equation}
In Fig. \ref{fig:4}, we plot the distributions for some of the clusters in our sample. 
We define the total likelihood function as
\begin{equation}
    \ln \mathcal{L}_{total}(e) = \sum_{i=1}^{3254} \ln \left(  \mathcal{P}(\mu,\sigma)*\mathcal{P}_{i}(e,\theta|S'(x,y)) \right)
    \label{eqn:19}
,\end{equation}
where $\mathcal{P}(\mu,\sigma)$ is the Gaussian model PDF, where $\mu$ is the mean of the normal distribution and $\sigma$ is the standard deviation, and $\mathcal{P}(e,\theta)$ are the posterior PDFs provided by MCMC in \citet[][]{Sanders_2025}, and $i$ is the $i^{th}$ cluster in our sample. To construct the probability distribution function of ellipticity, $\mathcal{P}_{obs}(e)$, for clusters in our sample, we maximized the total log-likelihood defined in equation \ref{eqn:19}, by varying the parameters $\mu$ and $\sigma$, and performed MCMC to obtain the parameters of the observational PDF. 

In Figs. \ref{fig:5} and \ref{fig:6}, we plot the observational PDF ($\mathcal{P}_{obs}(e)$) (black line), and the shaded region is the $1\sigma$ error. The mean ellipticity of our cluster sample is $e=0.79$. It is clear that most of the clusters in our sample deviate from spherical symmetry but are not extremely elongated.
\begin{table*}[]
\caption{Prior distributions of the samples for $l$ and $ w$.}    
\label{tab:1}     
\centering                         
\begin{tabular}{c c c c c c}   
\hline\hline           
Name         & Distribution of $l$ & Distribution of $w$ & Sample size & $\chi^2$  & final $\chi^2$  \\ \hline
Prior Sample 0    & $\mathcal{U}(1.0,5.0)$       & $\mathcal{U}(1,l)$& 500& 523.65 & 2.786   \\ 
Prior Sample 1    & $\mathcal{U}(1.25,2.5)$       & $\mathcal{U}(1,l)$& 200& 16.1 & 2.460   \\ 
Prior Sample 2 & $\mathcal{U}(2.5,5)$          & $\mathcal{U}(2,l)$ & 200   & 216.3 & 75.422  \\ 
Prior Sample 3 & $\mathcal{U}(1,3)$            & $\mathcal{U}(1,l)$  & 100  & 11.8 & 1.456  \\
Prior Sample 4 & $\mathcal{N}(1.5,0.2)$        & $\mathcal{N}(l,0.2)$ & 100  & 18.2 & 6.691\\ \hline
\hline                                   
\end{tabular}
\tablefoot{Prior distributions of the samples for $l$ and $ w$. Here, the $\mathcal{N}(\mu,\sigma)$ is the normal distribution with mean and standard deviation, and $\mathcal{U}(a, b)$ is the uniform distribution between $a$ and $b$. The last two columns list the $\chi^2$ statistic between the model PDF ($\mathcal{P}_{model}$) and the observational PDF ($\mathcal{P}_{obs}(e)$) before and after the Monte Carlo procedure.
}
\end{table*}
\subsection{Constraining the ellipticity PDF}
\label{Sec:3.2}
To find the most probable axis ratio pair(s) that can reconstruct our observational PDF, we first have to find the sample that contains the ratio pair(s). The model PDF (equation \ref{eqn:12}) constructed from points in the sample should resemble the observational PDF. To estimate the correct prior sample points (drawn from either a normal or a uniform distribution) that resemble our observation PDF, we sampled $l,w$ pairs from a distribution. We list the prior sample distributions in Table \ref{tab:1}. We simulated the conditional PDFs for all the $l,w$ pairs in the sample. We found the best-fit Gaussian to the conditional PDFs and summed the PDFs. We refer to this PDF as the model PDF (equation \ref{eqn:12}).

\begin{figure}[]
    \centering
    \includegraphics[width=\linewidth]{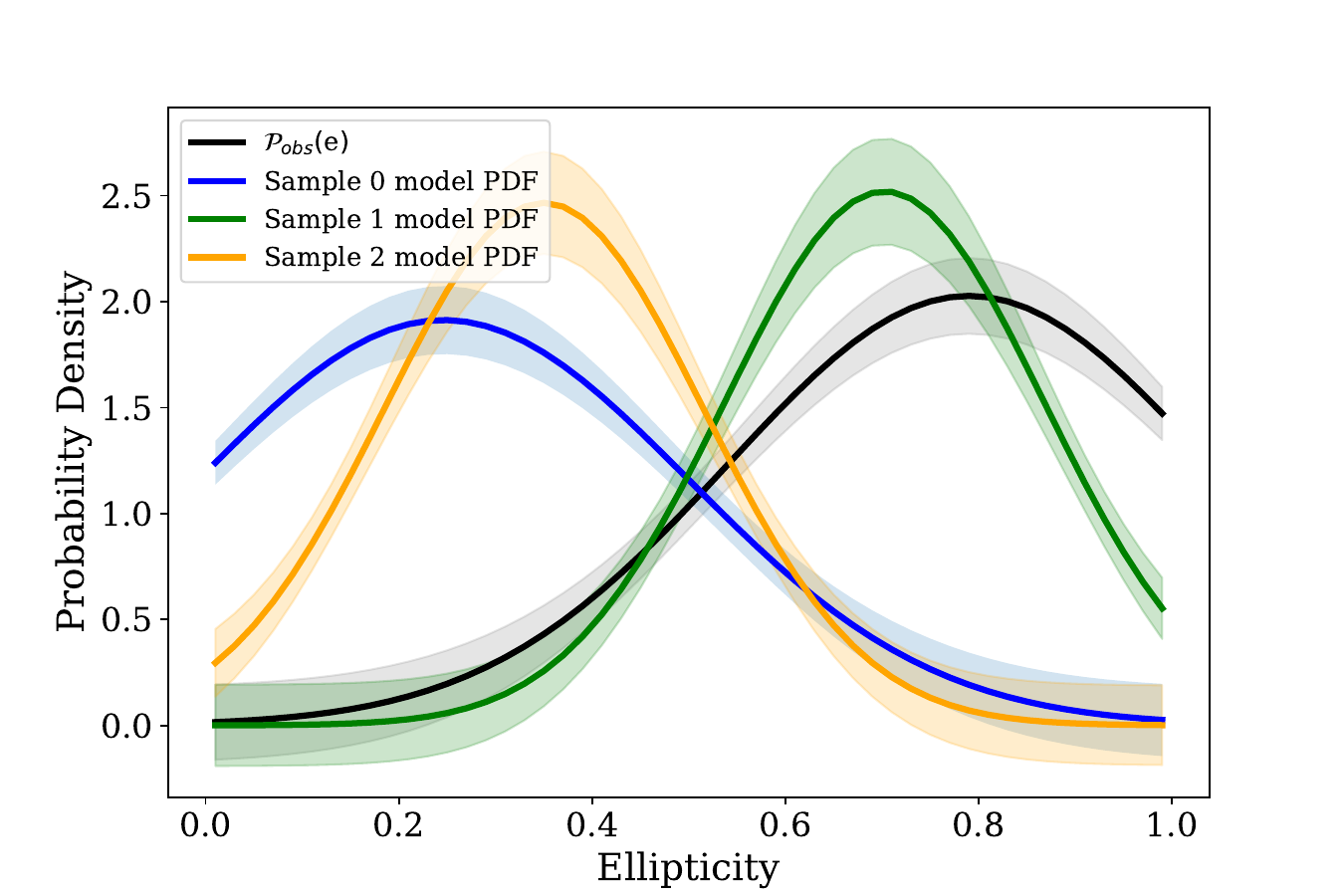}
    \caption{Model PDFs for Sample 0 (blue curve), Sample 1 (green curve), and Sample 2 (orange curve), sampled from distribution $l\in\mathcal{U}(1.0,5.0)$, $w\in\mathcal{U}(1,l)$ for sample 0, $l\in\mathcal{U}(1.25,2.5)$, $w\in\mathcal{U}(1,l)$ for sample 1, and $l\in\mathcal{U}(2.5,5)$, $w\in\mathcal{U}(1,l)$ for sample 2. The black curve represents the observational PDF.}
    \label{fig:5}
\end{figure}

In Table \ref{tab:1}, we tabulate the $\chi^2$ statistic between the observed PDF and the model PDF (equation \ref{eqn:12}) for the points drawn from the prior sample. We first used a non-informative, large uniform prior with 500 points. For an observational PDF described by a normal distribution, 500 points are sufficient and computationally efficient. While the initial $\chi^2$ for the complete prior sample with 500 points is high, the Monte-Carlo removes the unnecessary points in the sample, which is reflected by the final $\chi^2$. We plot the model PDF for prior sample 0 in figure \ref{fig:5} (blue curve). The model PDF differs from the observational PDF. The prior sample 0 model PDF peaks at $e \sim 0.23$. Since the prior sample contains many extreme shapes, the model PDF shifts to the left. Assuming that this model PDF contains the ratio pairs we are looking for, we can go ahead with the Monte-Carlo estimation. The final $\chi^2$ is still larger for prior sample 0 than for prior samples 1 and 3. This prior range does not contain the ratio pairs we are looking for, as it differs much from our observational PDF. Hence, a smaller, better-constrained prior is needed. 

The $\chi^2$ is the least for the prior samples 3 and 1, respectively. Both prior samples have a uniform distribution. Reducing the number of points in the prior sample for a similar sample distribution reduces the $\chi^2$, while decreasing the density of points in an $l,w$ bin. The most probable ratio pair(s) are present in the prior sample 3. We selected prior sample 3 over prior sample 1 due to the low $\chi^2$ estimates and smaller sample size. We plot the model PDF for sample 1 (green curve) and sample 2 (orange curve) in Figs. \ref{fig:5}. Sample 2 contains points from a uniform distribution, $l\in\mathcal{U}(2.5,5)$ and $w\in\mathcal{U}(2,l)$. The pairs from the sample represent extreme eccentric shapes. This is reflected in the model PDF. The peak of the model PDF for sample 2 distribution is closer to $e\sim0.35$. The final $\chi^2$ for this prior sample is much larger even after removing unnecessary points. Hence, this sample does not contain the ratio points, which are contained in our observational PDF.

The green PDF in Fig. \ref{fig:5} is similar to the observational PDF, but the probability density and the mean ellipticity are different. Hence, we did not select this sample. We took the points from sample 3 and performed a Monte-Carlo estimation for probability estimation.

\subsection{Monte-Carlo estimation}
\label{Sec:3.3}
We continued with the method described in section \ref{Sec:2.4}. We randomly removed a point from our sample and find the new model PDF. We accepted the removal if the $\chi^2$ was reduced; otherwise, we rejected the removal. The process stopped when the $\chi^2$ was no longer reduced and stabilized. We followed this procedure 15 times.
After each run, we were left with pairs of ratios that were essential to the new PDF (equation \ref{eqn:17}) and closely resembled our observational PDF. Removal of the points increased the probability of the remaining points.
\begin{figure}[]
    \centering
    \includegraphics[width=\linewidth]{ 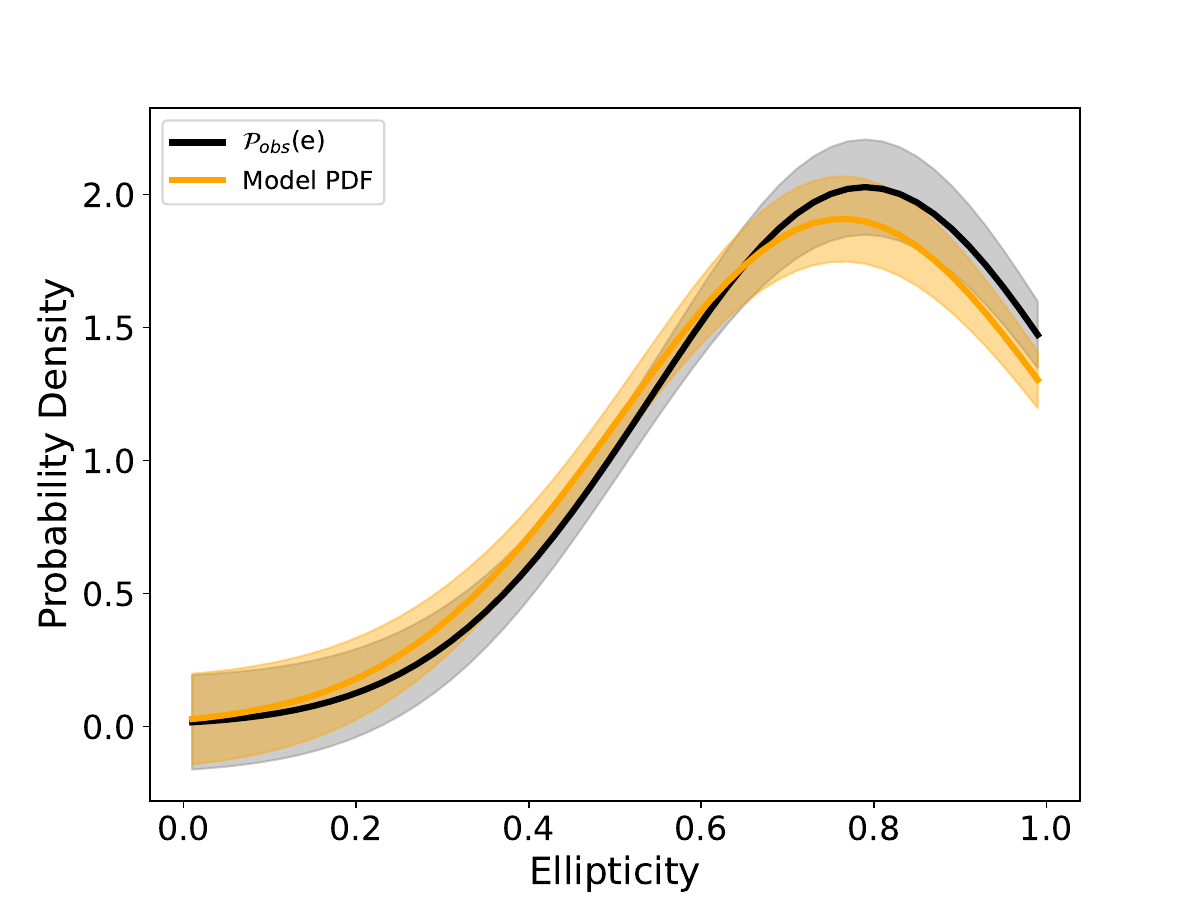}
    \caption{Model PDF for the points in reduced sample 3, i.e., after Monte-Carlo estimation procedure in section \ref{Sec:2.4}. The reduced sample was obtained by removing the points that are not essential to the model PDF. The model PDF closely follows the observational PDF (black curve), suggesting that the sample contains the correct pair of $l,w$, which reconstructs our observational PDF.}
    \label{fig:6}
\end{figure}
We plot the model PDF with the most probable points in Fig. \ref{fig:6}. Our model PDF closely resembles the observational PDF. Hence, our combination of the most probable points is correct, and the removed points do not necessarily contribute to the model PDF. We find the probability associated with the pairs of ratios by further binning the $l,w$ space. The concentration of the points in a bin increases the probability for that bin.

\section{Results}
\label{Sec4}
\begin{figure}[]
    \centering
    \includegraphics[width=\linewidth]{ 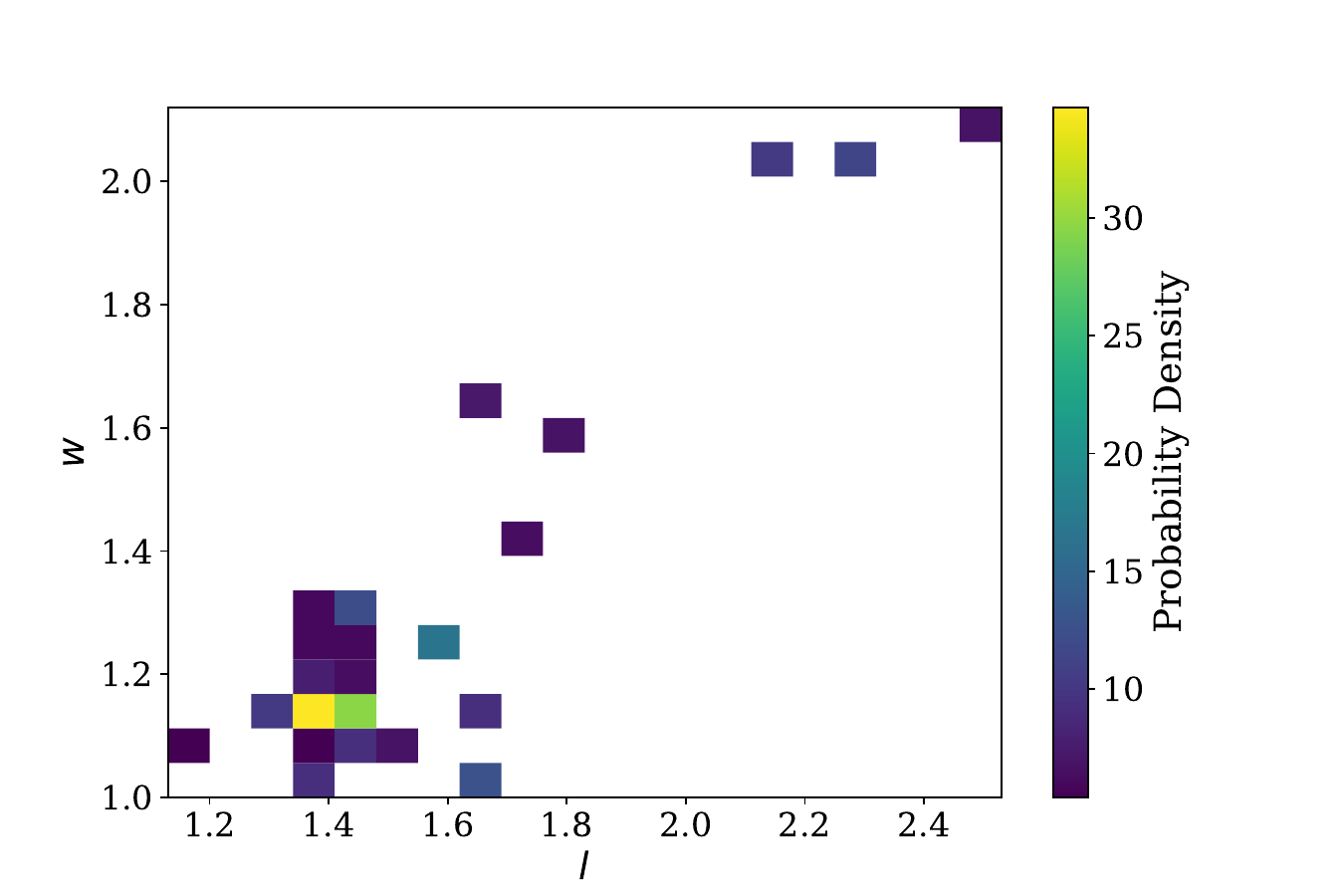}
    \caption{2D probability plot for the ratios in sample 3. The bar represents the probability density for the bins. The probability for the plot was evaluated by multiplying the probability by the width of the bins.}
    \label{fig:7}
\end{figure}

\begin{figure}
    \centering
    \includegraphics[width=\linewidth]{ 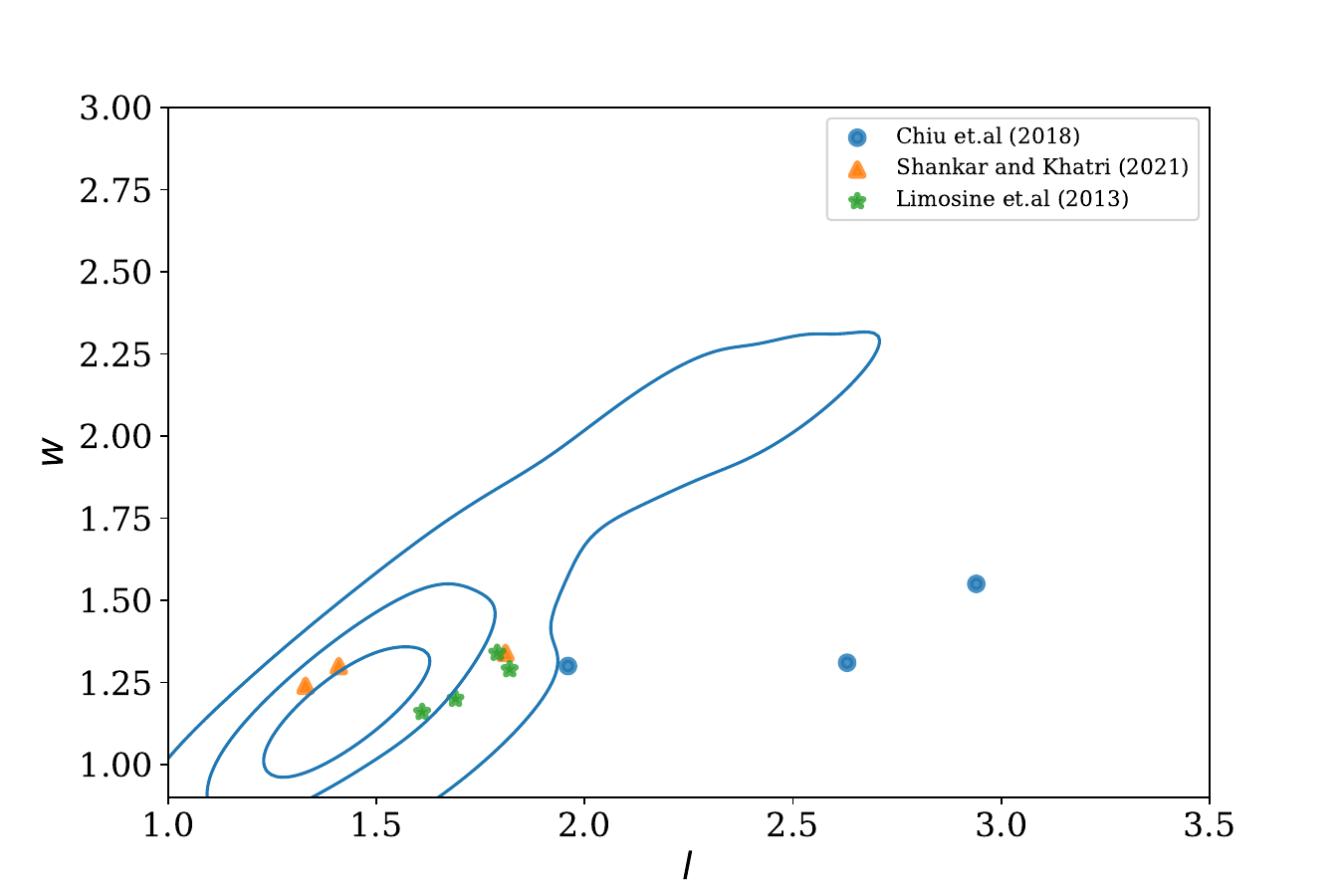}
    \caption{Comparison of our work with previous studies. The contours contain the 1$\sigma$, 2$\sigma$, and 3$\sigma$ probabilities of total probabilities estimated for Sample 3. The legend contains the studies with which we are comparing our results.}
    \label{fig:8}
\end{figure}

\begin{figure}[]
    \centering
    \includegraphics[width=\linewidth]{ 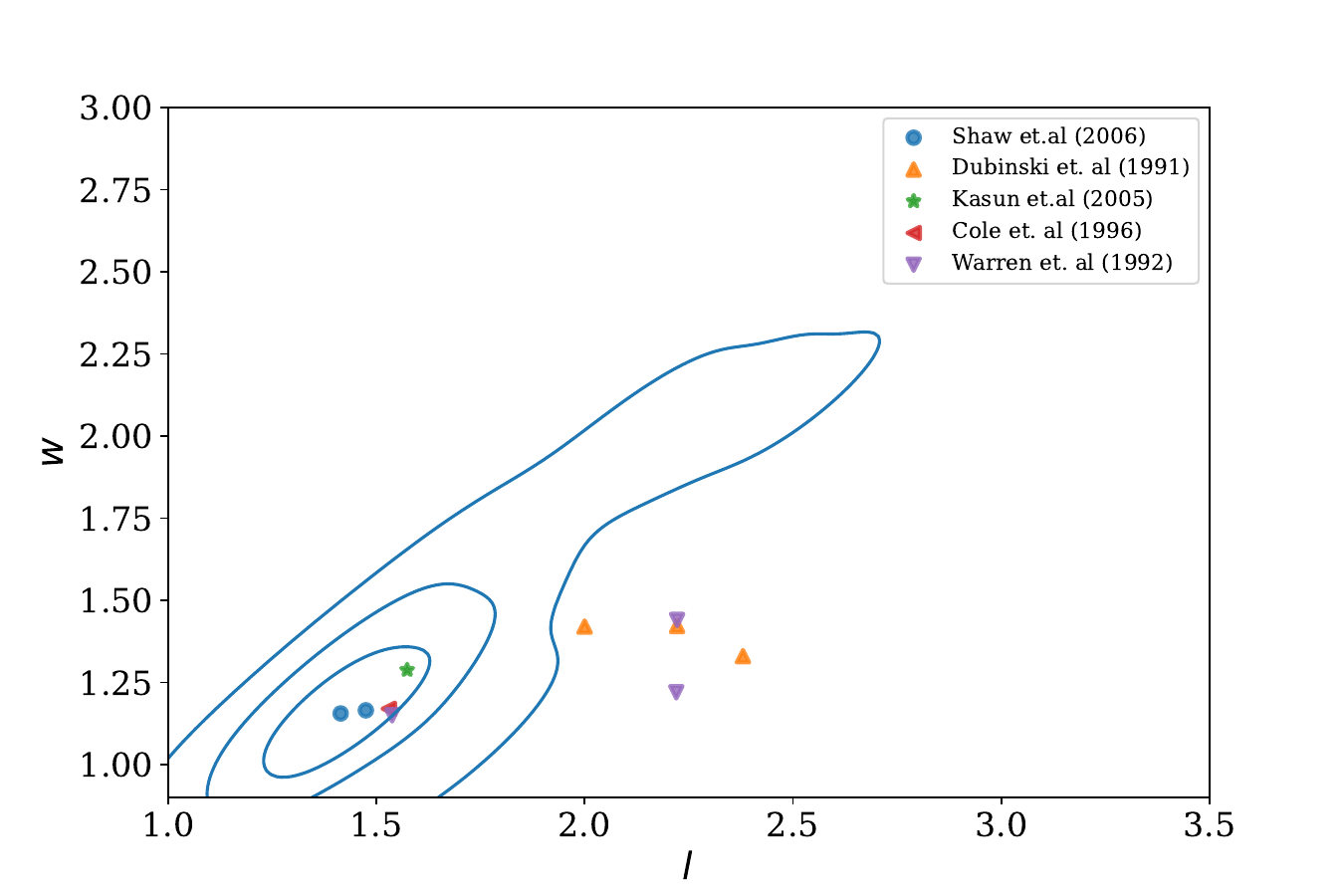}
    \caption{Comparison of our work with previous studies. The contours contain the 1$\sigma$, 2$\sigma$, and 3$\sigma$ probabilities of total probabilities estimated for Sample 3. The legend contains the studies with which we are comparing our results.}
    \label{fig:9}
\end{figure}

We started with a uniform distribution of 100 $(l,w)$ pairs in Sample 3. We removed the unnecessary points in this sample using Monte-Carlo estimation (section \ref{Sec:2.4}). We are now left with 39 points un-uniformly distributed in $(l,w)$ space. We plot the PDF using equation \ref{eqn:17} for the remaining points in Fig. \ref{fig:7}. The most probable points are clustered around $l=1.4$ and $w=1.1$ bins. We estimated the weighted mean and standard deviation for the most probable points in Sample 3 and found the 3D shape best describing the clusters in our eRASS1 subsample to be $(l,w)=(1.51 \pm 0.27,1.17 \pm 0.27)$($1\sigma$). This can be further confirmed by the fact that the $\chi^2$ for Sample 1 and Sample 4, (which contains ratio pairs from a normal distribution $l \in \mathcal{N}(1.5,0.2), \ w \in \mathcal{N}(l,0.2)$), are similar except for the prior distribution function. Hence, the mean of the normal distributions is close to the most probable point range, which explains the closeness of the $\chi^2$ estimate. 
In Fig. \ref{fig:7}, we show the most probable ratio bin with probability density, while in Figs. \ref{fig:8} and \ref{fig:9}, we compare the actual $(l,w)$ pairs with previous studies. The combination of the points in our sample reconstructs the model PDF; hence, all points remaining after Monte-Carlo estimation have a probability associated with them. 

We used both normal and uniform distributions for sampling the ratio pairs. The previous studies chose uniform distributions because the distribution gives equal probability to all the shapes, while the normal distribution gives preference to a specific shape. In our case, the normal distribution with $l\in\mathcal{N}(\mu,\sigma)=(1.5,0.2)$,$w\in\mathcal{N}(\mu,\sigma)=(l,0.2)$ approximately matches our observational PDF. With a better consensus about the most probable axial ratio, normal distributions might be used. 

We compare our results with previous studies in Fig. \ref{fig:8}. We plot the kernel density plot for the most probable points in sample 3. The blue contours represent the $1\sigma$, $2\sigma$, and $3\sigma$ levels of the total probability. As our study is motivated by the method of \citet{Shankar_2021}, it is a natural comparison. \citet{Limousin_2013} estimated the axial ratios for four strong lensing clusters (A1835, A383, A1689, and MACS 1423) using a joint dataset of X-ray, SZ, and lensing data. They report minor to major and intermediate to major axis ratios of dark matter halos, which we converted to $l,w$. \citet{Chiu_2018} analyzed the 20 clusters from the CLASH survey by combining the weak-lensing and strong lensing data. They were able to constrain only one axial ratio for 17 of their clusters without assuming any priors. We selected the clusters (A209, MACS J0329-0211, RX J1347-1145) for which both axial ratios are constrained. We would like to note that while \citet{Limousin_2013, Chiu_2018} have studied the shapes of individual clusters, we find the most probable ratio in a sample. The results of \citet{Limousin_2013} and \citet{Shankar_2021} lie within $3\sigma$ range of our total probability. For \citet{Chiu_2018}, our results lie outside the $3\sigma$ range. Our results may agree with \citet{Limousin_2013} better than \citet{Chiu_2018}, as they probe the shapes of halos using a joint dataset of X-ray, SZ, and lensing data, while \citet{Chiu_2018} probe the shapes by combining weak and strong lensing datasets.

In Fig. \ref{fig:9}, we plot the comparison of our results with N-body simulation results. Similar to \ref{fig:8}, the blue contours represent the $1\sigma$, $2\sigma$, and $3\sigma$ levels of the total probability. We assume HSE to compare the shapes of dark matter halos with our results. We compare our results with \citet{Dubinski_1991},\citet{Warren_1992}, \citet{Cole_1996},  \citet{Kasun_2005}, and \citet{Shaw_2006}. The N-body simulations probe the distribution of dark matter, while our results investigate the shape of the ICM. Our comparison is valid only under the assumption of HSE. While the assumption might not hold for individual cases, it works well as an approximation for a large sample of clusters. Our results slightly differ from \citet{Dubinski_1991} and \citet{Warren_1992}, as their study contains fewer particles and smaller dark matter halos than the other studies. The ellipticity of the halos with a smaller size (approximately kiloparsecs) might be different as the internal dynamics affect their shapes easily in the inner parts of the halos. Our results match well with \citet{Cole_1996, Kasun_2005}, and \citet{Shaw_2006}. We find that the clusters do not have spherical shapes, and clusters tend to be prolate. Our results agree with those of previous studies \citep{Schneider_2012, Parekh_2015}. 
\section{Conclusions}
\label{Sec5}
Galaxy clusters do not have completely spherical shapes. The assumptions of spherical shapes for clusters introduce biases in hydrostatic mass estimates of the clusters. These errors propagate further to cosmological studies. Estimation of the 3D shape of clusters requires multi-probe datasets. These observations are not available for all the clusters. Furthermore, observational and projection biases exist in shape measurements, leading to statistical estimatesz about shapes. In this study, we estimated the most probable 3D shape of galaxy clusters in eRASS1 using stereological methods, which can be used as a prior for future weak-lensing and cluster studies, as this is a much larger sample than previous studies. We were able to apply the method of \citet{Shankar_2021} despite the differences in the observational instruments and data reduction procedures.

Our observational PDF, $\mathcal{P}_{obs}(e)$, can be described by a normal distribution, $\mathcal{N}(\mu,\sigma)=(0.79,0.25)$. Next, we simulated conditional PDFs, $\mathcal{P}(e|l,w)$, for points drawn from a prior sample (either a normal distribution or a uniform distribution) to construct the model PDF. The prior sample, $l\in\mathcal{U}(1,3)$, $w\in\mathcal{U}(1,l)$, containing 100 points best matches our observational PDF.

We find the most probable shape of the clusters in our sample to be $(l,w)=(1.51 \pm 0.27,1.17 \pm 0.27)$. The clusters seem to prefer prolate shapes over oblate shapes. Our estimates agree with previously estimated cluster shapes and shapes obtained by the N-body simulations. Our results are in agreement with the X-ray studies of \citet{Shankar_2021} and the joint X-ray, SZ, and weak-lensing study of \citet{Limousin_2013}. Our results vary from those of \citet{Chiu_2018}, perhaps because they only use lensing data that probes the dark matter halo shapes, while our study and that of \citet{Limousin_2013} and \citet{Shankar_2021} probe the shapes of the ICM. Assuming that the HSE assumption holds for a large cluster sample, we compare our results with N-body simulation studies by \citet{Cole_1996},\citet{ Kasun_2005}, and \citet{Shaw_2006}. The results of these studies lie within a $1\sigma$ level of the total probability. Our results match well with the N-body simulation results.

Our results can be further improved by selecting a larger sample with better constraints, such as eRASS:5. This method can also be applied to optical and SZ cluster surveys to study the observational method biases.
\begin{acknowledgements}
We thank the anonymous referee for their helpful comments.
This work is based on data from eROSITA, the soft X-ray instrument aboard SRG, a joint Russian-German science mission supported by the Russian Space Agency (Roskosmos), in the interests of the Russian Academy of Sciences represented by its Space Research Institute (IKI), and the Deutsches Zentrum f\"ur Luft- und Raumfahrt (DLR). The SRG spacecraft was built by Lavochkin Association (NPOL) and its subcontractors, and is operated by NPOL with support from the Max Planck Institute for Extraterrestrial Physics (MPE). The development and construction of the eROSITA X-ray instrument was led by MPE, with contributions from the Dr. Karl Remeis Observatory Bamberg \& ECAP (FAU Erlangen-Nuernberg), the University of Hamburg Observatory, the Leibniz Institute for Astrophysics Potsdam (AIP), and the Institute for Astronomy and Astrophysics of the University of T\"ubingen, with the support of DLR and the Max Planck Society. The Argelander Institute for Astronomy of the University of Bonn and the Ludwig-Maximilians-Universit\"at Munich also participated in the science preparation for eROSITA. 

A.L. acknowledges the support from the National Natural Science Foundation of China (Grant No. 12588202). A.L. is supported by the China Manned Space Program with grant no. CMS-CSST-2025-A04.
\end{acknowledgements}

\bibliographystyle{aa} 
\bibliography{references.bib} 

\end{document}